\documentclass[aps,twocolumn,showpacs,byrevtex,prl,reprint, nofootinbib]{revtex4-1}
\usepackage{xspace}

\usepackage{epsfig}
\usepackage{dcolumn}
\usepackage{amssymb}   
\usepackage{amsmath}
\usepackage{mathrsfs}
\usepackage{bm}
\usepackage{color}
\usepackage{ulem}
\usepackage{graphicx}
\usepackage{subfigure}
\usepackage{pstricks}
\usepackage{pst-node}
\usepackage{rotating}
\usepackage{times}
\usepackage[english]{babel}
\addto{\captionsenglish}{%

}
\usepackage[colorlinks,linkcolor=blue, citecolor=blue]{hyperref}
\usepackage{lineno}
\newcommand{\ssb}{\Sigma^0\bar{\Sigma}^0}

\newcommand{\EE}{e^+e^-}

\newcommand{\jpsi}{J/\psi}
\newcommand{\ar}{\rightarrow}
\newcommand{\llb}{\Lambda\bar{\Lambda}}

\newcommand{\bfg}{\begin{figure}}
\newcommand{\efg}{\end{figure}}
\newcommand{\bitm}{\begin{itemize}}
\newcommand{\eitm}{\end{itemize}}
\newcommand{\bnum}{\begin{enumerate}}
\newcommand{\enum}{\end{enumerate}}
\newcommand{\btbl}{\begin{table*}}
\newcommand{\etbl}{\end{table*}}
\newcommand{\btbu}{\begin{tabular}}
\newcommand{\etbu}{\end{tabular}}
\newcommand{\bcl}{\begin{center}}
\newcommand{\ecl}{\end{center}}
\newcommand{\bbt}{\bibitem}
\newcommand{\beq}{\begin{equation}}
\newcommand{\eeq}{\end{equation}}
\newcommand{\beqr}{\begin{eqnarray}}
\newcommand{\eeqr}{\end{eqnarray}}

\begin{document}
\normalsize
\parskip=5pt plus 1pt minus 1pt
\title{\boldmath Measurement of $\Lambda$ baryon polarization in $\EE\ar\llb$ at $\sqrt{s} = 3.773$ GeV}
\author{
M.~Ablikim$^{1}$, M.~N.~Achasov$^{10,b}$, P.~Adlarson$^{68}$, S. ~Ahmed$^{14}$, M.~Albrecht$^{4}$, R.~Aliberti$^{28}$, A.~Amoroso$^{67A,67C}$, M.~R.~An$^{32}$, Q.~An$^{64,50}$, X.~H.~Bai$^{58}$
% ...
\\
\vspace{0.2cm}
(BESIII Collaboration)\\
\vspace{0.2cm} {\it
$^{1}$ Institute of High Energy Physics, Beijing 100049, People's Republic of China\\
% ...
$^{4}$ Bochum Ruhr-University, D-44780 Bochum, Germany\\
% ...
$^{10}$ G.I. Budker Institute of Nuclear Physics SB RAS (BINP), Novosibirsk 630090, Russia\\
% ...
$^{14}$ Helmholtz Institute Mainz, Staudinger Weg 18, D-55099 Mainz, Germany\\
% ...
$^{28}$ Johannes Gutenberg University of Mainz, Johann-Joachim-Becher-Weg 45, D-55099 Mainz, Germany\\
% ...
$^{32}$ Liaoning Normal University, Dalian 116029, People's Republic of China\\
% ...
$^{50}$ State Key Laboratory of Particle Detection and Electronics, Beijing 100049, Hefei 230026, People's Republic of China\\
% ...
$^{58}$ University of Jinan, Jinan 250022, People's Republic of China\\
% ...
$^{64}$ University of Science and Technology of China, Hefei 230026, People's Republic of China\\
% ...
$^{67}$ University of Turin and INFN, (A)University of Turin, I-10125, Turin, Italy; (B)University of Eastern Piedmont, I-15121, Alessandria, Italy; (C)INFN, I-10125, Turin, Italy\\
$^{68}$ Uppsala University, Box 516, SE-75120 Uppsala, Sweden\\
% ...
$^{b}$ Also at the Novosibirsk State University, Novosibirsk, 630090, Russia\\
% ...
}
}

\date{\today}

\begin{abstract}
  Using a data sample of $\psi(3770)$ events collected with the BESIII detector at BEPCII corresponding to an integrated luminosity of 2.9 fb$^{-1}$, we report a measurement of $\Lambda$ spin polarization in $\EE\ar\llb$ at $\sqrt{s} = 3.773$ GeV. 
The significance of polarization is found to be 2$\sigma$ including the systematic uncertainty, which implies a zero phase between the transition amplitudes of the $\Lambda\bar{\Lambda}$ helicity states. This phase can be interpreted in terms of psionic form factors, and is determined to be $\Delta\Phi^{\Psi}$ = $\Phi^{\Psi}_{E} - \Phi^{\Psi}_{M}$ = $(71^{+66}_{-46}$ $\pm$ 5)$^{\circ}$. Similarly, the ratio between the form factors is found to be $R^{\psi}$ = $|G^{\Psi}_{E}/G^{\Psi}_{M}|$ = $0.48^{+0.21}_{-0.35}$ $\pm$ 0.03. The first uncertainties are statistical and
the second systematic.
\end{abstract}
\maketitle

The study of baryon pair production in $\EE$ annihilation provides  an ideal system to probe the structure of baryons. The importance of baryon structure was pointed out as early as 1960~\cite{Cabibbo:1961sz}, but until recently, most experimental investigations have focused on protons and neutrons (for a review, see Ref. \cite{perdrisat}). 
It is straightforward to access the spacelike electromagnetic structure of protons , which is quantified in terms of electromagnetic form factors (EMFFs), through elastic electron-proton scattering.
This is not the case for unstable baryons with finite lifetime which cannot be used in such scattering experiments. Instead, $\EE$ processes allow access to timelike EMFFs.The timelike form factors are related to more intuitive spacelike quantities such as charge and magnetization densities by dispersion relations~\cite{Belushkin:2006qa}. 
At $\EE$ center-of-mass (CM) energies that do not overlap with vector resonances,
baryon pair production is dominated by one-photon exchange. The pair production of spin-1/2 baryons can then be parametrized by the electric form factor $G_{E}$ and the magnetic form factor $G_{M}$ which are analytic functions of the momentum transfer squared, $q^2$. In the timelike region, where $q^2 > 0$, the EMFFs are complex and have a relative phase $\Delta\Phi = \Phi_{E} - \Phi_{M}$. The phase is a reflection of interfering production amplitudes, and has a polarizing effect on the final state even if the initial state is unpolarized \cite{theorypolarization}. This 
provides a handle to study the asymptotic properties of the EMFFs: at large $|q^2|$, the space-like and time-like EMFFs should converge to the same value. For protons, the onset of this scale can be studied by measuring spacelike and timelike EMFFs. Ground-state baryons, however, have an advantage that compensates for their inaccessibility in the space-like region: their weak parity-violating decay gives straightforward access to the polarization.

Until recently, experimental data on baryon EMFFs was very limited. 
The {\it BABAR} collaboration reported the first results on the cross sections for processes $\EE\ar\llb$, $\ssb$, $\Lambda\Sigma^{0}$ by using the initial state radiation (ISR) technique. They measured the effective form factor, which is related to both the cross section assuming one-photon exchange and to the modulus of the ratio $R = |G_{E}/G_{M}|$. 
 However, because of the limited data sample, the experiment reported a limited on the relative phase between the $\Lambda$ electric and magnetic form factors~\cite{Aubert:2007uf}.
Subsequently, the CLEO collaboration measured the cross sections of baryon pairs ($p$, $\Lambda$, $\Sigma^0$, $\Xi^-$ and $\Omega^-$) at the charmonium resonances~\cite{Dobbs:2014ifa, Dobbs:2017}. Their conclusions regarding EMFFs and diquark correlations rely on the assumption that one-photon exchange dominates the production process and that decaying charmonia contributions are negligible~\cite{Dobbs:2014ifa,Dobbs:2017}. 
In the vicinity of $\psi(3770)$ peak,  the $\EE\ar\llb$ is still dominated by virtual photon, although the statistical significance is about 4$\sigma$ for $\psi(3770)\ar\llb$~\cite{bes3llbar}. 
The BESIII collaboration has also studied the $\Lambda$ EMFFs with data taken at several CM energy points from 2.2324 to 3.08 GeV~\cite{Ablikim:2017pyl, bes3prl}.

In the vicinity of vector charmonia, the spin formalism of Ref.~\cite{Faldt:2017kgy} is still valid, except that the amplitudes no longer represent electromagnetic form factors but instead hadronic or \textit{psionic} form factors, $G^{\Psi}_E$ and $G^{\Psi}_M$.
Polarization effects in $\EE$ collisions were neglected in previous studies~\cite{Ablikim:2016iym, Ablikim:2016iym-01, Ablikim:2016iym-02, Ablikim:2016iym-03, Ablikim:2016iym-04, Ablikim:2016iym-05, Ablikim:2016iym-06, Wang:2018kdh, Wang:2021lfq}.
Recently, the $\Sigma^+$ baryon polarization was studied by the BESIII collaboration in $\EE\ar\jpsi, \psi(3686)\ar \Sigma^+\bar{\Sigma}^-$ processes~\cite{yanliang}. The results do not only reveal a nonzero relative psionic phase, but also that the phase changes sign at the $\psi(3686)$ mass with respect to the value measured at the $J/\psi$ resonance. 
Subsequently, the $\Lambda$ polarization and $\Lambda\bar{\Lambda}$ entanglement were studied in the $\EE\ar\jpsi\ar\llb$ process by the BESIII collaboration~\cite{Ablikim:2018zay}, and study of $\EE\ar\psi(3686)\ar\llb$ is underway.
 The measurement in Ref.~\cite{bes3prl} is performed in the low-energy, off-resonance region where one-photon exchange is a valid assumption, in contrast to the measurement at the $J/\psi$ resonance \cite{Ablikim:2018zay} where production through $J/\psi$ should be completely dominating. The energy point at 3.773 GeV is interesting in this regard since we learned from Ref.~\cite{bes3llbar} that the production occurs through an interplay of one-photon and $\psi(3770)$ exchange, i.e. its amplitudes represent EM-\textit{psionic} form factor.
The production at 3.773 GeV is dominated by different mechanisms compared to the reactions in Refs.~\cite{Ablikim:2018zay,bes3prl}, and the study in this work will be complementary to them.
The very large data set corresponding to an integrated luminosity of 2.9 fb$^{-1}$, collected at the CM energy of 3.773 GeV~\cite{Ablikim:2014gna}
with the BESIII detector~\cite{Wang:2007tv} at BEPCII~\cite{BESIII}, enables such a study for the first time. 

\begin{figure}[!htbp]
\includegraphics[width=0.4\textwidth]{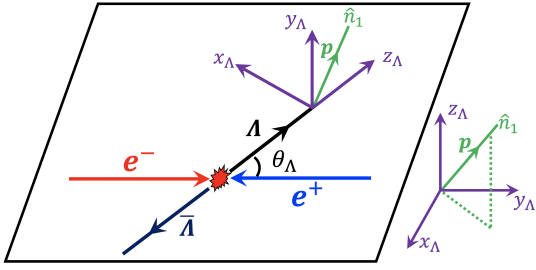}
\caption{
Definition of the coordinate system used to describe the $\EE\ar\llb\ar p\bar{p}\pi^{+}\pi^{-}$ reaction. The $\Lambda$ particle is emitted along the $z_{\Lambda}$ axis direction, and the $\bar\Lambda$ in the opposite direction. The $y_{\Lambda}$ axis is perpendicular to the plane of $\Lambda$ and $e^{-}$, and the $x_{\Lambda}$ axis is defined by a right-hand coordinate system. 
The $\Lambda$ decay product, the proton, is measured in this coordinate system.}
\label{fig:helicity_frame}
\end{figure}

The $\EE\ar\gamma/\Psi\ar\llb\ar p\bar{p}\pi^{+}\pi^{-}$ process
can be fully described by the $\Lambda$ scattering angle in the CM system of the reaction, $\theta_{\Lambda}$, and the $p(\bar{p})$ direction in the rest frame of its parent particle, $\boldsymbol{\hat{n}_{1}}$($\boldsymbol{\hat{n}_{2}}$).
We use a right-handed system for each baryon decay as defined in Fig.~\ref{fig:helicity_frame}, with the $z$-axis is defined along the $\Lambda$ momentum $\textbf{p}_{\Lambda} = - \textbf{p}_{\bar{\Lambda}^{-}} = \textbf{p}$ in the CM system. The $y$-axis is taken as the normal to the scattering plane, $\textbf{k}_{e^{-}} \times \textbf{p}_{\Lambda}$, where $\textbf{k}_{e^{-}} = - \textbf{k}_{e^{+}} = \textbf{k}$ is the electron beam momentum in the CM system. 
For the determination of the psionic form factor ratio $R^{\Psi}$ and phase $\Delta\Phi^{\Psi}$, the angular distribution $\alpha_{\Psi}$ (but not its absolute normalization) is of interest. In Ref.~\cite{Faldt:2017kgy}, the joint decay angular distribution of the processes $\EE\ar\llb\ar p\bar{p}\pi^{+}\pi^{-}$ is expressed in terms of the phase $\Delta\Phi^{\Psi}$ and the angular distribution parameter

\begin{align}
 \label{eq:tangles:abcd}
{\cal{W}}(\boldsymbol{\xi}) = & 1 + \alpha_{\Psi}\cos^{2}\theta_{\Lambda} \nonumber\\
+ &\alpha_{\Lambda}\alpha_{\bar\Lambda}[\sin^{2}\theta_{\Lambda}(n_{1, x}n_{2, x} -\alpha_{\Psi}n_{1, y}n_{2, y}) \nonumber \\
+ &(\cos^{2}\theta_{\Lambda} +\alpha_{\Psi})n_{1, z}n_{2, z}] \\
+&\alpha_{\Lambda}\alpha_{\bar\Lambda}[\sqrt{1-\alpha_{\Psi}^{2}}\cos\Delta(\Phi^{\Psi})\sin\theta_{\Lambda}\cos\theta_{\Lambda}
(n_{1, x}n_{2, z} \nonumber \\
+ & n_{1, z}n_{2, x})]\nonumber \\
+ &\sqrt{1-\alpha_{\Psi}^{2}}\sin\Delta(\Phi^{\Psi})\sin\theta_{\Lambda}\cos\theta_{\Lambda}(\alpha_{\Lambda}n_{1, y} +\alpha_{\bar\Lambda}n_{1, y}), \nonumber
\end{align}
where 
$\alpha_{\Lambda(\bar\Lambda)}$ denotes the decay asymmetry of the $\Lambda(\bar\Lambda)\ar p\pi^{-}(\bar{p}\pi^{+})$ decay.
 The scattering angle distribution parameter $\alpha_{\Psi}$, is related to the ratio $R^{\Psi}$ by
\begin{equation}
R^{\Psi}=\sqrt{ \frac{\tau(1-\alpha_{\Psi})}{1+\alpha_{\Psi}}},
\end{equation}
where $\tau = s/4m^{2}_{\Lambda}$, and $s$ is the square
of the CM energy.
If the initial state is unpolarized, and the production process is either strong or electromagnetic and hence parity conserving, then a nonzero polarization is only possible in the transverse, or $y$, direction. The polarization is given by
\begin{equation}
P_y=\frac{\sqrt{1-\alpha_{\Psi}^2}\sin\theta_{\Lambda}\cos\theta_{\Lambda}}{1+\alpha_{\Psi}\cos^2\theta_{\Lambda}}\sin(\Delta\Phi^{\Psi}).
\label{eq:pol}
\end{equation}

Full reconstructed $\EE\ar\llb$ events with $\Lambda\ar p\pi^-$ and $\bar\Lambda\ar\bar{p}\pi^+$ are selected for further analysis. 
To determine the detection efficiency for the decay $\EE\ar\llb$ at $\sqrt{s} = 3.773$ GeV, 200 million Monte Carlo (MC) events are generated  according
to a phase space model corresponding to $\alpha_{\Psi} = 0$ using the \textsc{kkmc} generator~\cite{kkmc, kkmc-01}, which
includes the initial state radiation (ISR) effect.  The $\Lambda(\bar\Lambda)$ decays into $p\pi^{-}(\bar{p}\pi^{+})$ are simulated using \textsc{evtgen}~\cite{evt2, evt2-01}. The response of the BESIII detector is modeled with MC simulations
using a framework based on \textsc{geant}{\footnotesize 4}~\cite{geant4,geant4-01}.

Charged tracks are required to be reconstructed in the multilayer drift chamber (MDC) within its angular coverage: $|\cos\theta|<0.93$, where $\theta$ is the polar angle with respect to the $e^{+}$ beam direction in the laboratory system.  
Events with two negatively charged tracks and two positively charged tracks
are kept for further analysis.

To reconstruct $\Lambda(\bar\Lambda)$ candidates, a secondary vertex fit~\cite{XUM} is applied to all combinations of one positively charged track and one negatively charged track. From all combinations, the one with the minimum value
of $\sqrt{|M_{p\pi^{-}}-m_{\Lambda}|^{2} + |M_{\bar{p}\pi^{+}}-m_{\bar\Lambda}|^{2}}$ is selected. Here, $M_{p\pi^{-}(\bar{p}\pi^{+})}$ is the invariant mass of the $p\pi^{-}(\bar{p}\pi^{+})$ pair, and $m_{\Lambda(\bar\Lambda)}$ is the known mass of $\Lambda(\bar\Lambda)$ taken from the PDG~\cite{PDG2020}. The combinations with $\chi^{2} < 500$ with 3 degrees of freedom are kept for further analysis. To further suppress background from non-$\Lambda$ events, the $\Lambda$ decay length is required to be greater than zero, where the negative decay lengths are due to the detector resolution.  

\begin{figure}[!htbp]
\includegraphics[width=0.45\textwidth]{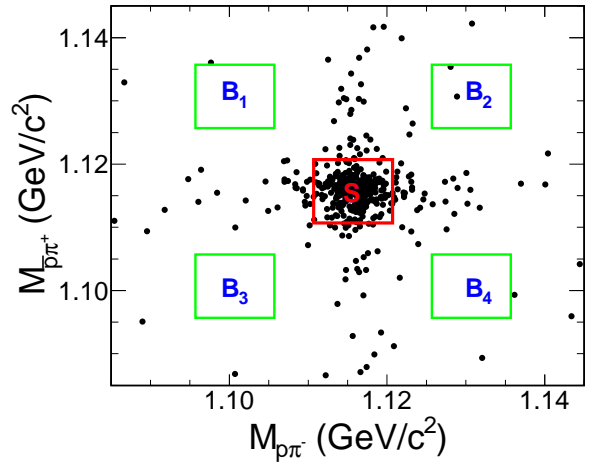}
\caption{Two-dimensional distribution of $M_{\bar{p}\pi^{+}}$ versus $M_{p\pi^{-}}$ for data, where the red solid box indicates the signal region, the green dashed boxes show the selected background regions.
}
\label{scatter_plot::llb}
\end{figure}

To suppress further background contributions and improve the mass resolution, a four-constraint (4C) kinematic fit imposing energy-momentum conservation from the initial $\EE$ to the final $\llb$ state is applied for all $\llb$ hypotheses after the $\Lambda(\bar\Lambda)$
reconstruction, combined with the requirement of $\chi^{2}_{\rm 4C} < 200$. 
Figure~\ref{scatter_plot::llb} shows the distribution of
$M_{\bar{p}\pi^{+}}$ versus $M_{p\pi^{-}}$.
A clear accumulation of events around $\Lambda$ mass can be seen.
The $p\pi^{-}(\bar{p}\pi^{+})$
invariant mass, $M_{p\pi^{-}(\bar{p}\pi^{+})}$ is required to be within 5 MeV/$c^{2}$ of the known
$\Lambda(\bar\Lambda)$ mass. The signal region marked by $S$ in Fig.~\ref{scatter_plot::llb} is determined with the figure of merit
$\frac{S}{\sqrt{S + B}}$ based on the MC simulation, where $S$ is the
number of signal MC events and $B$ is the number of the background events expected from simulation of generic $\EE\ar$ hadron events.
After applying the event selection criteria to the
data, the remaining background contamination in this analysis comes mainly from non-$\Lambda(\bar\Lambda)$ events, such as $\EE\ar\pi^{+}\pi^{-}p\bar{p}$. The number of background events is estimated using the corner method, {\it i.e.} $\sum^{4}_{i=1}B_{i}/4$ for $M_{p\pi^{-}}$ and $M_{\bar{p}\pi^{+}}$ windows, where $i$ runs over the four regions shown in Fig.~\ref{scatter_plot::llb}, {\it i.e.} 
  \begin{itemize}
\item $B_{1}: [1.0957, 1.1057]$ and $[1.1257, 1.1357]$ GeV/$c^{2}$,
\item $B_{2}: [1.1257, 1.1357]$ and $[1.1257, 1.1357]$ GeV/$c^{2}$,
\item $B_{3}: [1.0957, 1.1057]$ and $[1.0957, 1.1057]$ GeV/$c^{2}$,
\item $B_{4}: [1.1257, 1.1357]$ and $[1.0957, 1.1057]$ GeV/$c^{2}$.
  \end{itemize}
The size of the final data sample is 262 events with
a statistical uncertainty of 16 events. The estimated background from the aforementioned corner method is $\sim$2 events.
 This implies a background level of $\approx$0.5\%, which is a negligible contamination of the signal.

\begin{figure}[!htbp]
\includegraphics[width=0.23\textwidth]{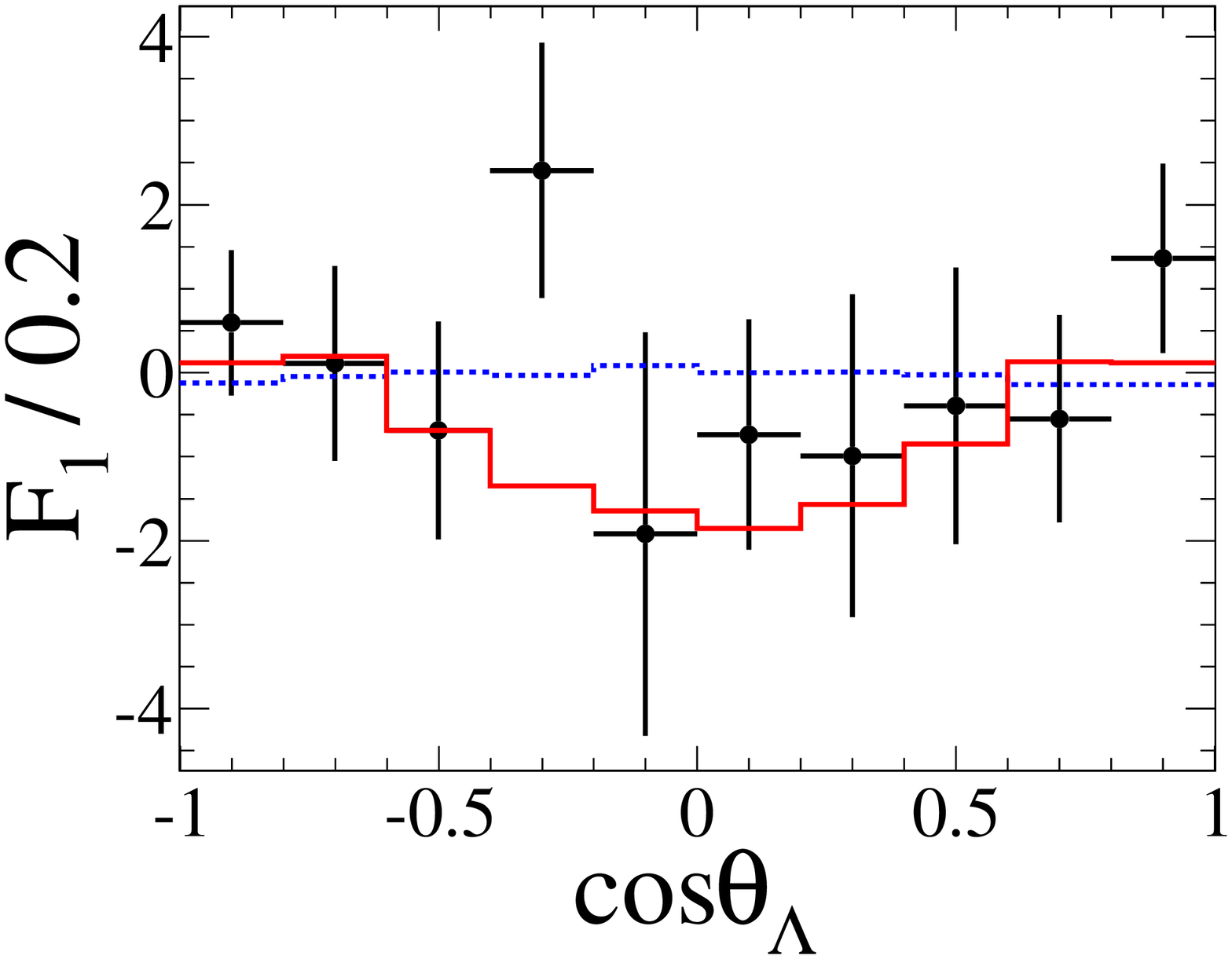}
\includegraphics[width=0.23\textwidth]{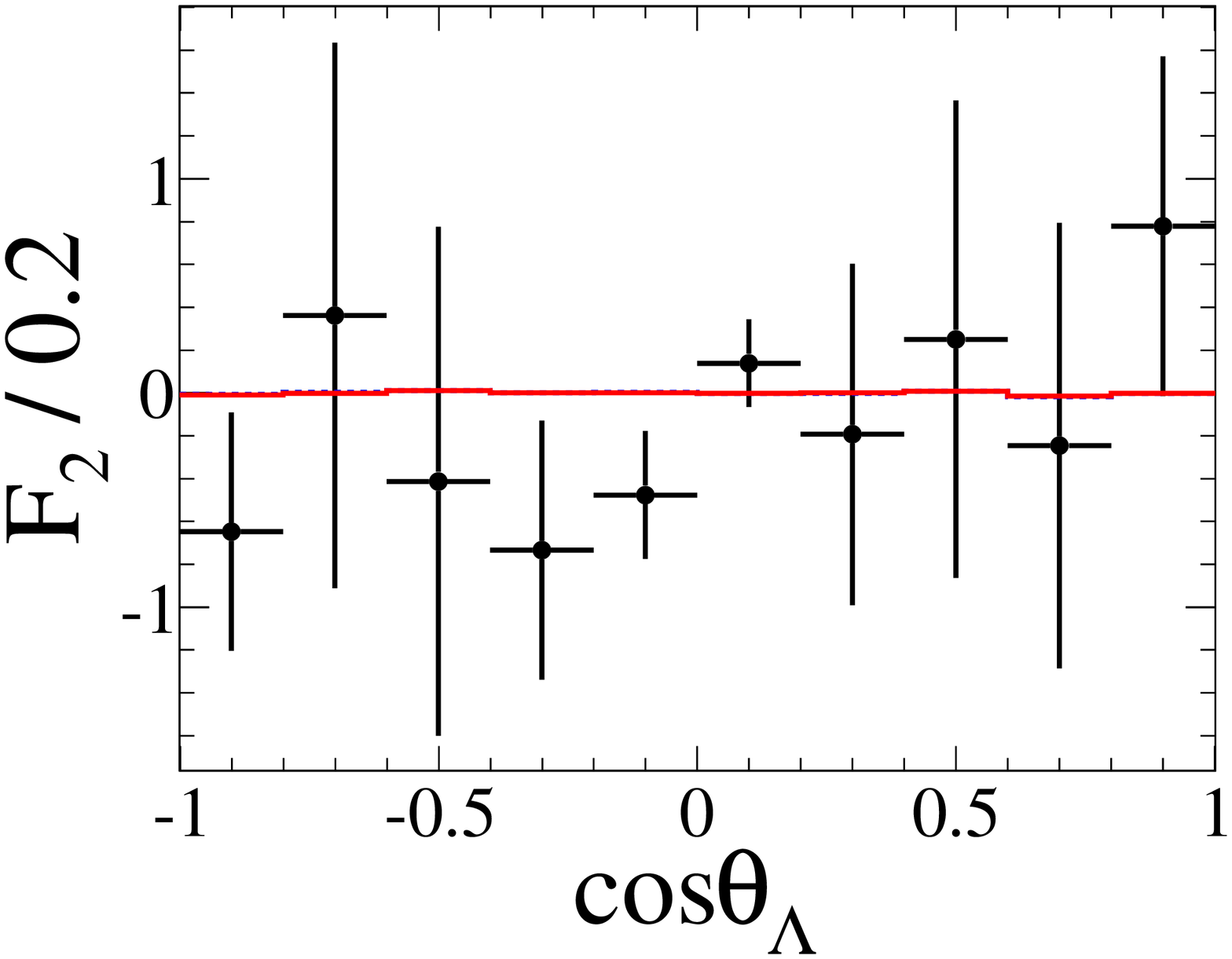}\\
\includegraphics[width=0.23\textwidth]{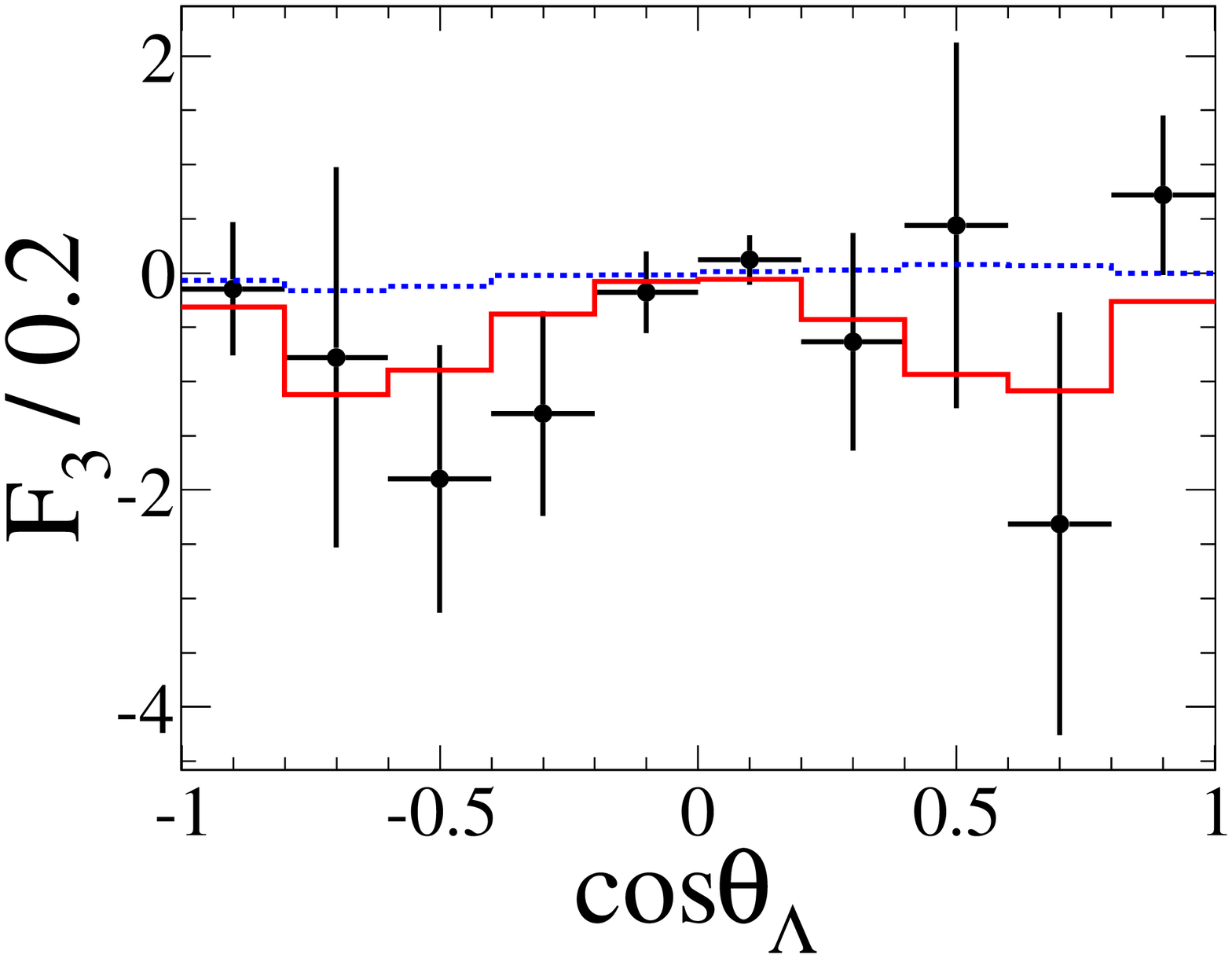}
\includegraphics[width=0.23\textwidth]{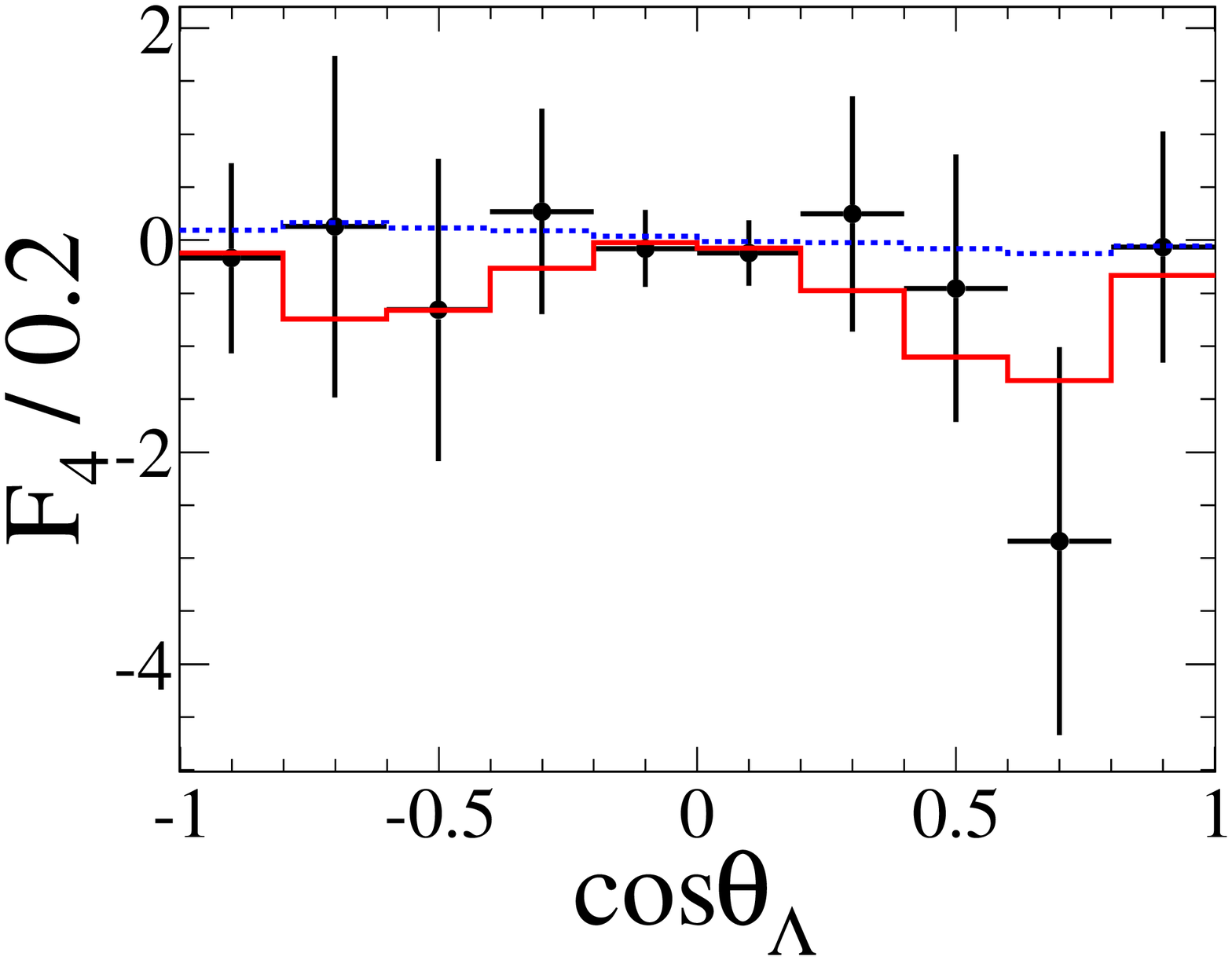}\\
\includegraphics[width=0.23\textwidth]{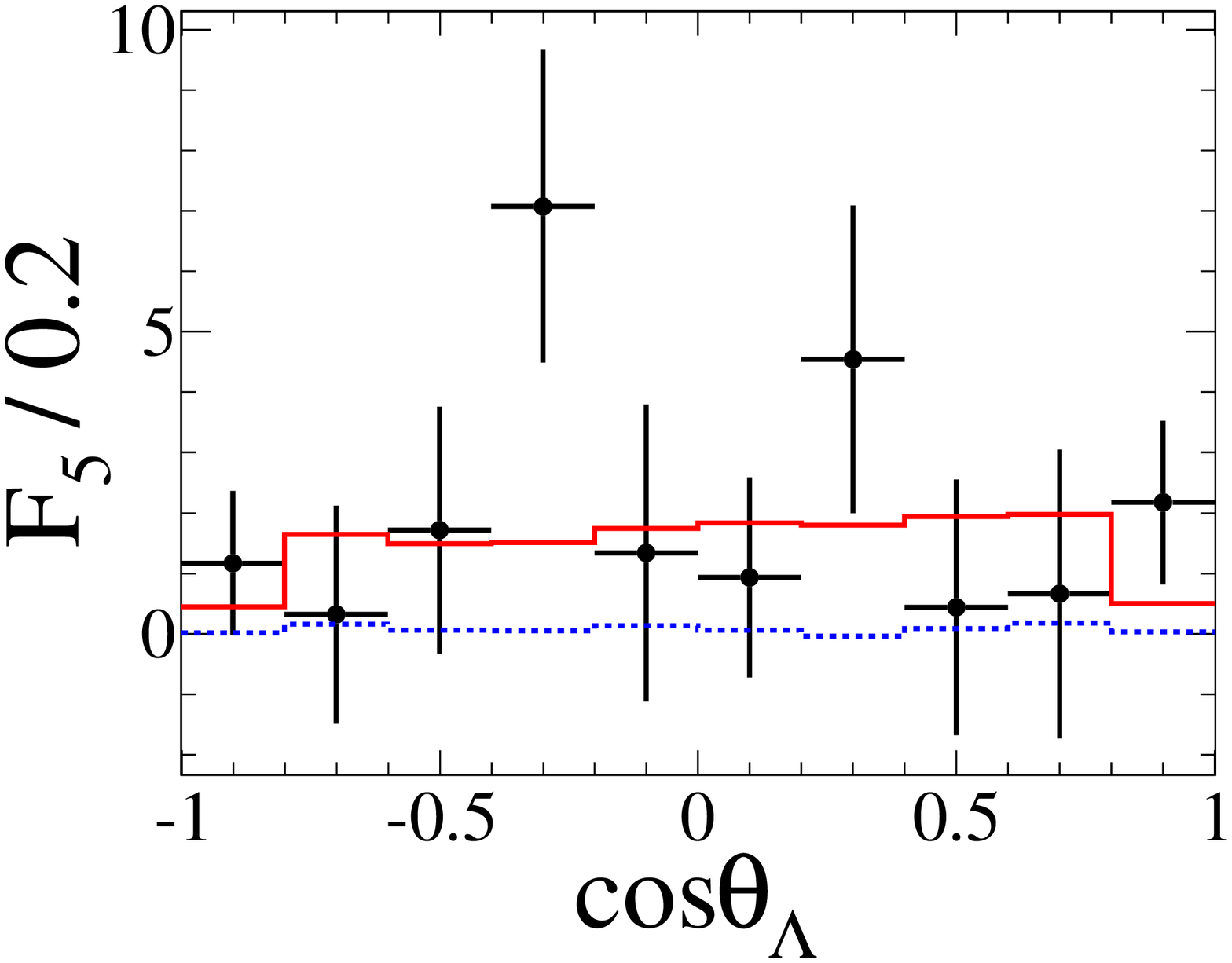}
\includegraphics[width=0.23\textwidth]{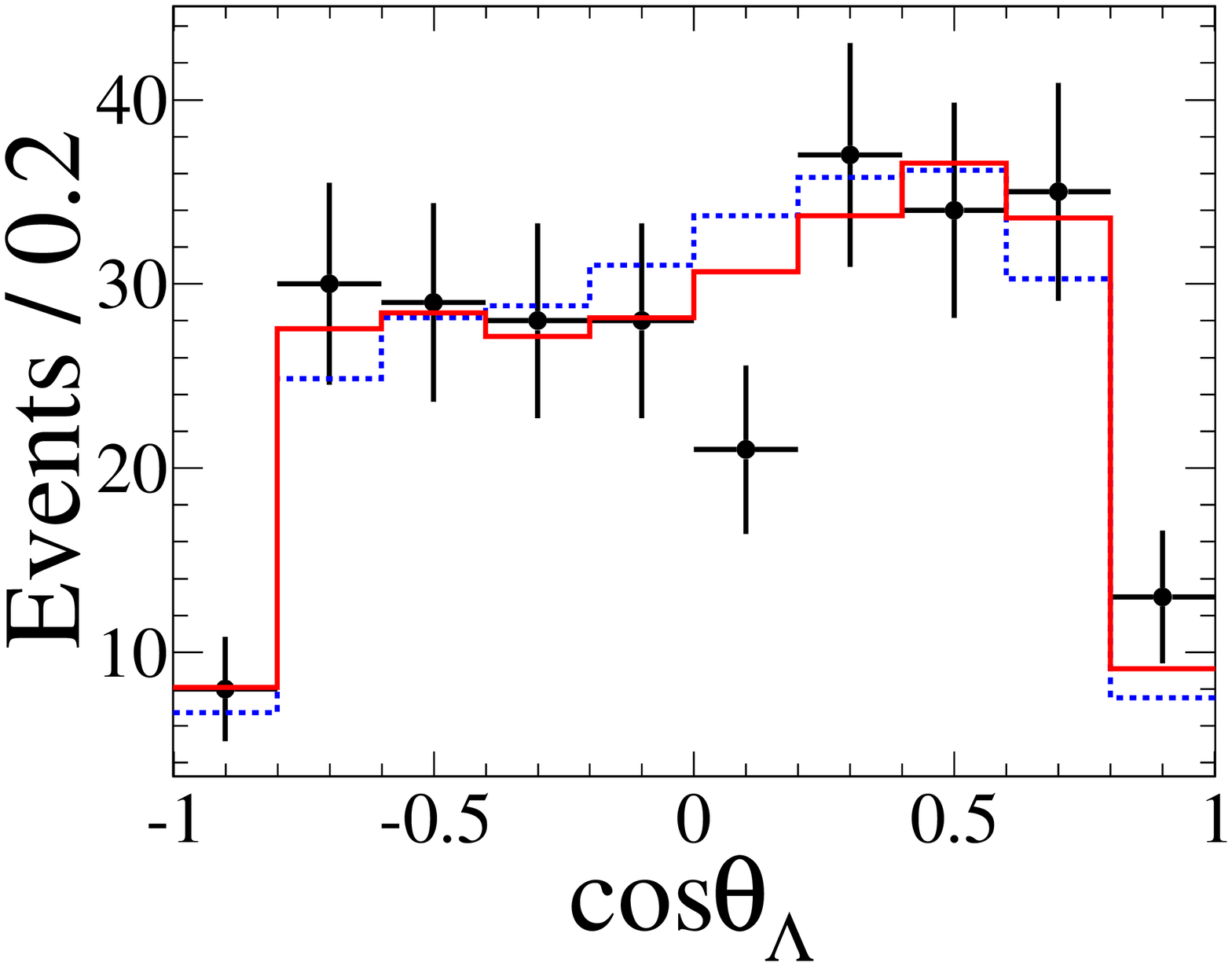}
\caption{
Distributions of ${\cal{F}}_{k} (k = 1, 2, ...5)$ moments with respect to the $\cos\theta_{\Lambda}$ and the $\cos\theta_{\Lambda}$  distribution (bottom right). The dots with error bars are data, and the red lines are the fit results. 
The blue dashed line represents the distribution of the simulated events evenly distributed in phase space, without polarization.}
\label{scatter_plot::llb:projections}
\end{figure}

To determine the set of $\Lambda$ spin polarization parameters $\boldsymbol{\Omega} = \{\Delta\Phi^{\Psi}, \alpha_{\Psi}\}$,  an unbinned maximum likelihood fit is performed to extract the decay parameters, where the decay parameters $\alpha_{\Lambda/\bar\Lambda}$ are fixed to the value 0.754 obtained from the average in Ref.~\cite{Ablikim:2018zay} assuming $CP$ conservation.
In the fit,  the likelihood function $\mathscr{L}$ is constructed from the probability density function (PDF), ${\cal{P}}({\boldsymbol{\xi}}_i)$, for an event $i$ characterized by the measured angles $\boldsymbol{\xi}_i$

\begin{equation}
\mathscr{L} = \prod_{i=1}^N {\cal{P}}({\boldsymbol{\xi}}_i, {\boldsymbol{\Omega}}) = \prod_{i=1}^N {\cal{C}}{\cal{W}}({\boldsymbol{\xi}}_i, {\boldsymbol{\Omega}})\epsilon(\boldsymbol{\xi}_i), 
 \end{equation}
where $N$ is the number of events in the signal region. The joint angular distribution ${\cal{W}}({\boldsymbol{\xi}}_i, {\boldsymbol{\Omega}})$ is given in Eq.~(\ref{eq:tangles:abcd}), and $\epsilon(\boldsymbol{\xi}_i)$ is the detection efficiency.
The normalization factor $\mathcal{C}=\frac{1}{N_\mathrm{MC}}\sum_{j=1}^{N_\mathrm{MC}} {\cal{W}}({\boldsymbol{\xi}}^{j}, {\boldsymbol{\Omega}})$ is given by the sum of the corresponding amplitude $\cal{W}$ using the accepted MC events $N_\mathrm{MC}$, and the difference between data and MC simulation is taken into account. 
The objective function minimization defined as  
\begin{equation}
\mathit{S} = -\mathrm{ln}\mathscr{L}_{data} + \mathrm{ln}\mathscr{L}_{bg},
\end{equation}
is performed with the MINUIT package from the CERN library~\cite{James:1975dr}.
Here, $\mathscr{L}_{data}$ is the likelihood function of events selected in the signal region, and $\mathscr{L}_{bg}$ is the likelihood function of background events determined in the sideband regions. 
Figure~\ref{scatter_plot::llb:projections} shows fitted distributions of the five moments 
${\cal{F}}_{k} (k = 1, 2, ...5)$ with respect to the $\cos\theta_{\Lambda}$
defined in Eq.~(\ref{eq:tangles}) and the $\cos\theta_{\Lambda}$ distribution.
\begin{align}
 \label{eq:tangles}
{\cal{F}}_1 = &\sum_i^{N(m)}(\sin^2\!\theta_{\Lambda} n^i_{1, x}n^i_{2, x} + \cos^2\!\theta_{\Lambda} n^i_{1, z}n^i_{2, z}),\nonumber\\
{\cal{F}}_2 = &\sum_i^{N(m)}\sin\!\theta_{\Lambda}\cos\!\theta_{\Lambda} (n^i_{1, x}n^i_{2, z} + n^i_{1, z}n^i_{2, x}),\nonumber\\
{\cal{F}}_3 = &\sum_i^{N(m)}\sin\!\theta_{\Lambda}\cos\!\theta_{\Lambda} n^i_{1, y},\\
{\cal{F}}_4 = &\sum_i^{N(m)}\sin\!\theta_{\Lambda}\cos\!\theta_{\Lambda} n^i_{2, y},\nonumber\\
{\cal{F}}_5 = &\sum_i^{N(m)}(n^i_{1, z}n^i_{2, z} - \sin^2\!\theta_{\Lambda} n^i_{1, y}n^i_{2, y}).\nonumber
\end{align}
The moments are calculated for 10 intervals  in $\cos\theta_{\Lambda}$, $N(m)$ is the number of events in the $m^{th}$ $\cos\theta_{\Lambda}$ interval.
The numerical fit results, with asymmetric uncertainties, are summarized in Table~\ref{table:sum_decay}.

\begin{figure}[!htbp]
 \includegraphics[width=0.45\textwidth]{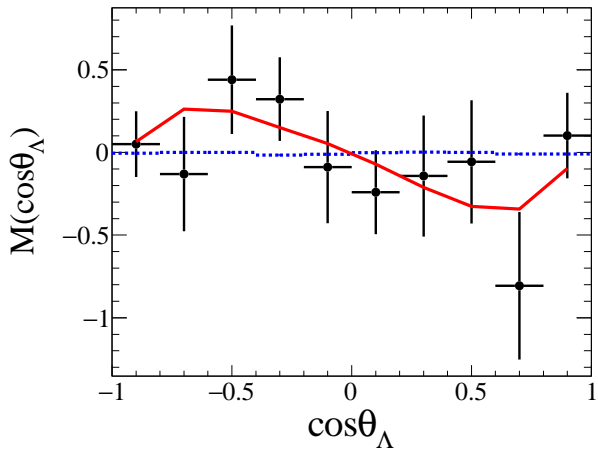}
\caption{The moments $M(\cos\theta_{\Lambda})$ as a function of $\cos\theta_{\Lambda}$ for the $\EE\ar\llb$ process at $\sqrt{s} = 3.773$ GeV. Points with error bars are data, the red solid line is the fit result of the global fit based on Eq.~(\ref{moment}), and the blue dashed line represents the distribution without polarization from simulated events, evenly distributed in the phase space.}
\label{scatter_plot::llb:polarization}
\end{figure}

Figure~\ref{scatter_plot::llb:polarization} shows the result of the fit in the $P_{y}$ distribution, which is consistent with
the behavior of Eq.~(\ref{eq:pol}) as compared to the data.
The significance of the polarization signal is found to be 2$\sigma$ considering the systematic uncertainties,
and is estimated by comparing the likelihoods of the baseline fit and the one assuming no polarization. Therefore, with the current data sample, the relative phase is compatible with zero. The effect by fixing the decay parameters $\alpha_{\Lambda/\bar\Lambda}$ values
is estimated conservatively by varying the parameters by one standard derivation, and the combination with the smallest significance is adopted.
The moment given by
\begin{equation}\label{moment}
 M(\cos\theta_{\Lambda}) = \frac{m}{N}\sum_i^{N(m)}(n^{i}_{1,y} - n^{i}_{2,y}),
\end{equation}
is related to the polarization, and calculated for $m = 10$
intervals in $\cos\theta_{\Lambda}$. Here, $N$ is the total number of events in the data sample, and $N(\cos\theta_{\Lambda})$ is the number of events in the $\cos\theta_{\Lambda}$ intervals.
 In the limit of $CP$ conservation, $\alpha_{\Lambda} = - \alpha_{\bar\Lambda}$, and the expected angular dependence is $M(\cos\theta_{\Lambda})\sim\sqrt{1-\alpha_{\Psi}^{2}}\alpha_{\bar\Lambda}\!\sin\Delta\Phi^{\Psi}\cos\theta_{\Lambda}\sin\theta_{\Lambda}$ as shown in Fig.~\ref{scatter_plot::llb:polarization} according to Eq.~(\ref{eq:tangles:abcd}). 

Systematic uncertainties on the measurement of the
$\Lambda$ baryon polarization in the $\EE\ar\llb$ process arise due to the fit method, the requirements on the $p\pi^{-}$ mass window, the kinematic fit
and the decay parameters of $\Lambda\ar p\pi$. 
To validate the reliability of the fit results, an input and output check based on 500 pseudoexperiments is performed with the helicity amplitude formula in Ref.~\cite{Ablikim:2018zay}. The polarization parameters measured in this analysis are used as input in the formula, and the number of events in each generated MC sample is the same as in the data sample.
The differences between the input and output results are taken as the systematic uncertainty caused by the fitting method.
The uncertainty due to the $\Lambda$ reconstruction including the tracking, the requirements on the mass window and on the decay length of 
the $\Lambda$ is determined from a control sample of $\jpsi\ar\llb$ events. The uncertainty is found to be negligible.
The systematic uncertainty due to the kinematic fit is estimated  with the same control sample, $\jpsi\ar\llb$, with and without a helix correction~\cite{yanliang}, and it is found to negligible. The uncertainties from the decay parameters of $\Lambda\ar p\pi^{-}$, $\alpha_{\Lambda/\bar\Lambda}$, are estimated by varying the baseline value, obtained from averaging the results in Ref.~\cite{Ablikim:2018zay}, by $\pm$1$\sigma$. The largest difference in the result is taken as the systematic uncertainty.
Assuming all sources are independent,
the total systematic uncertainties on the measurement of 
the polarization parameters  for the $\EE\ar\llb\ar p\bar{p}\pi^{+}\pi^{-}$ process are determined by
the square root of the quadratic sum of these sources as listed
in Table~\ref{tab:Xi:Rec:eff:polarization}.

\begin{table}[!hbpt]
\begin{center}
\caption{The absolute systematic uncertainty on the measurements of the $\Lambda$ baryon polarization parameters.}
\begin{tabular}{cccc} \hline\hline
Source                                                   &$\alpha_{\Psi}$ &$\Delta\Phi$ (rad) &$R^{\Psi}$ \\ \hline
Fit method                                             &0.02&0.08 &0.03\\
Decay parameter                                     &0.01&0.02 &0.02\\
Total                                                       &0.02&0.08 &0.04\\ \hline\hline
\end{tabular}
\label{tab:Xi:Rec:eff:polarization}
\end{center}
\end{table}

In summary, using a data sample corresponding to an integrated luminosity of 2.9 fb$^{-1}$ collected with the BESIII detector at BEPCII, we report the measurement of $\Lambda$ spin polarization in $\EE\ar\llb$ process at $\sqrt{s} = 3.773$ GeV.
 The relative EM-$psionic$ form factor phase, the ratio of the form factors and the angular distribution parameter are determined to be
$\Delta\Phi^{\Psi}$ = $(71^{+66}_{-46}$ $\pm$ 5)$^{\circ}$, $R^{\Psi}$ = $0.48^{+0.06}_{-0.10}$ $\pm$ 0.01,  and $\alpha_{\Psi} =$ $0.85^{+0.21}_{-0.35}$ $\pm$ 0.03, respectively, where the first uncertainties are statistical and the second ones systematic.
The measured phase differs from zero with a significance of 2$\sigma$ including the systematic uncertainties. 
 A comparison between this and previous measurements in  $\EE\ar\llb$~\cite{Ablikim:2017pyl} at 2.396 GeV and $\jpsi\ar\llb$~\cite{Ablikim:2018zay} is shown in Table~\ref{table:sum_decay}. Within the relatively large uncertainties, the value of the phase does not vary considerably between the different energy points. 
But the $\alpha_{\Psi}$ values are obviously different for three energy points. It is a hint of different production mechanisms among them.
More data samples at different energy points are needed for a detailed understanding of the phase dependence on the momentum transfer squared, $q^2$. Spin-1/2 baryons produced in a baryon-antibaryon pair can have either the same or opposite helicity.
A nonvanishing phase $\Delta\Phi$  between the transition amplitudes of these helicity states implies that not only the $s$-wave but also the $d$-wave amplitude contribute to the $\llb$ production. This manifests itself in a polarized final state.
The uncertainty of the measured  $\Delta\Phi$ value at 3.773 GeV is large due to the limited data sample size. In order to better understand the underlying $\llb$ production mechanism and the structure of the $\Lambda$ baryon, a larger data sample at this, and at other energy points, such as the $J/\psi$ peak, would be illuminating~\cite{BESIII:2020nme}.
\begin{table*}[htbp]
\caption{\label{table:sum_decay} The measured parameters compared with other measurements. The first uncertainty is statistical and the second one is systematic. The symbol "..." indicates that the parameter was not measured as part of this work in question.
}
\begin{ruledtabular}
\begin{tabular}{lccc} 
Parameters    &This work    &$\jpsi\ar\llb$~\cite{Ablikim:2018zay}    &$e^{+}e^{-}\ar\llb$ ($\sqrt{s}$ = 2.396 GeV)~\cite{bes3prl} \\ \hline
$\alpha_{\Psi}$ &$0.85^{+0.12}_{-0.20}$ $\pm$ 0.02 &0.461 $\pm$ 0.006 $\pm$ 0.007 &0.12 $\pm$ 0.14 $\pm$ 0.02 \\ 
$\Delta\Phi^{\Psi}$ ($^\circ$) &$71^{+66}_{-46}$  $\pm$ 5 &42.4 $\pm$ 0.6 $\pm$ 0.5  &37 $\pm$ 12 $\pm$ 6  \\ 
$R^{\Psi}$            &$0.48^{+0.21}_{-0.35}$ $\pm$ 0.03 &0.843 $\pm$ 0.006 $\pm$ 0.007 &0.96 $\pm$ 0.14 $\pm$ 0.02\\ 
\end{tabular}
\end{ruledtabular}
\end{table*}
\section{Acknowledgement}
\label{sec:acknowledgement}
The BESIII collaboration thanks the staff of BEPCII and the IHEP computing center for their strong support. This work is supported in part by National Key R\&D Program of China under Contracts No. 2020YFA0406400, No. 2020YFA0406300; National Natural Science Foundation of China (NSFC) under Contracts No. 12075107, No. 11905236, No. 12047501, No. 11625523, No. 11635010, No. 11735014, No. 11822506, No. 11835012, No. 11935015, No. 11935016, No. 11935018, No. 11961141012, No. 12022510, No. 12025502, No. 12035009, No. 12035013, No. 12061131003; the Chinese Academy of Sciences (CAS) Large-Scale Scientific Facility Program; Joint Large-Scale Scientific Facility Funds of the NSFC and CAS under Contracts No. U1732263, No. U1832207; CAS Key Research Program of Frontier Sciences under Contract No. QYZDJ-SSW-SLH040; 100 Talents Program of CAS; 
Fundamental Research Funds for the Central Universities under Grants No. lzujbky-2021-sp24;
INPAC and Shanghai Key Laboratory for Particle Physics and Cosmology; ERC under Contract No. 758462; European Union Horizon 2020 research and innovation programme under Contract No. Marie Sklodowska-Curie grant agreement No 894790; German Research Foundation DFG under Contracts Nos. 443159800, Collaborative Research Center CRC 1044, GRK 214; Istituto Nazionale di Fisica Nucleare, Italy; Ministry of Development of Turkey under Contract No. DPT2006K-120470; National Science and Technology fund; Olle Engkvist Foundation under Contract No. 200-0605; STFC (United Kingdom); The Knut and Alice Wallenberg Foundation (Sweden) under Contract No. 2016.0157; The Royal Society, UK under Contracts Nos. DH140054, DH160214; The Swedish Research Council; U. S. Department of Energy under Contracts No. DE-FG02-05ER41374, No. DE-SC-0012069.


\begin{thebibliography}{**}
\bibitem{Cabibbo:1961sz} N.~Cabibbo and R.~Gatto, 
\href{https://journals.aps.org/pr/abstract/10.1103/PhysRev.124.1577}{Phys.\ Rev.\  {\bf 124}, 1577 (1961).}
\bibitem{perdrisat} C.~F.~Perdrisat and V.~Punjabi and M. Vanderhaeghen, 
\href{https://www.sciencedirect.com/science/article/pii/S0146641007000610}{Prog.\ Part.\ Nucl.\ Phys.\ \textbf{59}, 694 (2007)}.
\bibitem{Belushkin:2006qa} M.~A.~Belushkin, H.~W.~Hammer and U.~G.~Meissner, 
\href{https://journals.aps.org/prc/abstract/10.1103/PhysRevC.75.035202}{Phys.\ Rev.\ C {\bf 75}, 035202 (2007).}
\bibitem{theorypolarization} A. Z.~Dubnickova {\it et al.}, 
\href{https://link.springer.com/article/10.1007%2FBF02731012}{Nuovo Cimento Soc. Ital. Fis.  {\bf 109A}, 241 (1996).}
\bibitem{Aubert:2007uf}  B.~Aubert {\it et al.} ({\it BABAR} Collaboration),  
\href{https://journals.aps.org/prd/abstract/10.1103/PhysRevD.75.092006}{Phys.\ Rev.\ D {\bf 76}, 092006 (2007).}
\bibitem{Dobbs:2014ifa} S.~Dobbs, A.~Tomaradze, T.~Xiao, K.~K.~Seth, and G.~Bonvicini, 
\href{https://www.sciencedirect.com/science/article/pii/S0370269314007485?via%3Dihub}{Phys.\ Lett.\ B {\bf 739}, 90 (2014).} 
\bibitem{Dobbs:2017} S.~Dobbs, K.~K.~Seth, A.~Tomaradze, T.~Xiao, and G.~Bonvicini, 
\href{https://journals.aps.org/prd/abstract/10.1103/PhysRevD.96.092004}{Phys.\ Rev.\ D {\bf 96}, 092004 (2017).}
\bibitem{bes3llbar} M.~Ablikim \textit{et al.} (BESIII Collaboration), 
\href{https://journals.aps.org/prd/abstract/10.1103/PhysRevD.104.L091104}{Phys. Rev. D \textbf{104}, L091104 (2021).}

\bibitem{Ablikim:2017pyl}  M.~Ablikim {\it et al.} (BESIII Collaboration),
  \href{https://journals.aps.org/prd/abstract/10.1103/PhysRevD.97.032013}{Phys.\ Rev.\ D {\bf 97}, 032013 (2018).}
  

  
  \bibitem{bes3prl} M.~Ablikim {\it et al.} (BESIII Collaboration), 
  \href{https://journals.aps.org/prl/abstract/10.1103/PhysRevLett.123.122003}{Phys. Rev. Lett. \textbf{123}, 122003 (2019).}
\bibitem{Faldt:2017kgy} G.~F\"aldt and A.~Kupsc,
\href{https://www.sciencedirect.com/science/article/pii/S0370269317304719?via%3Dihub}{Phys.\ Lett.\ B {\bf 772}, 16 (2017).}  


\bibitem{Ablikim:2016iym} M.~Ablikim {\it et al.} (BESIII Collaboration), 
\href{http://journals.aps.org/prd/abstract/10.1103/PhysRevD.87.032007}{Phys.\ Rev.\ D {\bf 87}, 032007 (2013).}
\bibitem{Ablikim:2016iym-01} M.~Ablikim {\it et al.} (BESIII Collaboration), 
\href{http://journals.aps.org/prd/abstract/10.1103/PhysRevD.93.072003}{Phys.\ Rev.\ D {\bf 93}, 072003 (2016).}
\bibitem{Ablikim:2016iym-02} M.~Ablikim {\it et al.} (BESIII Collaboration), 
\href{https://www.sciencedirect.com/science/article/pii/S0370269317303222?via%3Dihub}{Phys.\ Lett.\ B {\bf 770}, 217 (2017).}
\bibitem{Ablikim:2016iym-03} M.~Ablikim {\it et al.} (BESIII Collaboration), 
\href{https://journals.aps.org/prd/abstract/10.1103/PhysRevD.100.051101}{Phys.\ Rev.\ D {\bf 100}, 051101 (2019).}
\bibitem{Wang:2018kdh} X.~F.~Wang, B.~Li, Y.~N.~Gao and X.~C.~Lou,  
\href{https://www.sciencedirect.com/science/article/pii/S0550321319300689?via%3Dihub}{Nucl.\ Phys. {\bf B941}, 861 (2019).}
\bibitem{Ablikim:2016iym-04} M.~Ablikim {\it et al.} (BESIII Collaboration), 
\href{https://journals.aps.org/prl/abstract/10.1103/PhysRevLett.124.032002}{Phys.\ Rev.\ Lett.\ {\bf124}, 032002 (2020).}
\bibitem{Ablikim:2016iym-05} M.~Ablikim {\it et al.} (BESIII Collaboration), 
\href{https://journals.aps.org/prd/abstract/10.1103/PhysRevD.103.012005}{Phys. Rev. D \textbf{103}, 012005 (2021).}
\bibitem{Ablikim:2016iym-06} M.~Ablikim {\it et al.} (BESIII Collaboration), 
\href{https://www.sciencedirect.com/science/article/pii/S0370269321004974?via%3Dihub}{Phys.\ Lett.\ B {\bf 820}, 136557 (2021).}
\bibitem{Wang:2021lfq}
X.~F.~Wang, \href{https://pos.sissa.it/385/026}{PoS \textbf{Proc. Sci. CHARM2020} (2021) 026.}


 \bibitem{Ablikim:2018zay} M.~Ablikim {\it et al.} (BESIII Collaboration), 
 \href{https://www.nature.com/articles/s41567-019-0494-8/}{Nat. Phys.\  {\bf 15}, 631 (2019).}


    \bibitem{yanliang} M.~Ablikim {\it et al.} (BESIII Collaboration), 
\href{https://journals.aps.org/prl/abstract/10.1103/PhysRevLett.125.052004}{Phys.\ Rev.\ Lett.\  {\bf 125}, 052004 (2020).}
    
 \bibitem{Ablikim:2014gna} M.~Ablikim {\it et al.}  (BESIII Collaboration), 
 \href{https://iopscience.iop.org/article/10.1088/1674-1137/37/12/123001}{Chin.\ Phys.\ C {\bf 37}, 123001 (2013).}
 
\bbt{Wang:2007tv} Y.~F.~Wang, 
\href{https://www.worldscientific.com/doi/abs/10.1142/S0217751X06034513}{Int.\ J.\ Mod.\ Phys.\ A {\bf 21}, 5371 (2006).}
\bbt{BESIII} M.~Ablikim {\it et al.}  (BESIII Collaboration), 
\href{https://www.sciencedirect.com/science/article/pii/S0168900209023870?via%3Dihub}{Nucl.\ Instrum.\ Methods Phys. Res., Sect.\ A {\bf 614}, 345 (2010).}
\bbt{kkmc} S.~Jadach, B.~F.~L.~Ward and Z.~Was, 
\href{https://www.sciencedirect.com/science/article/pii/S0010465500000485?via%3Dihub}{Comput.\ Phys.\ Commun.\  {\bf 130}, 260 (2000).}
\bbt{kkmc-01} S.~Jadach, B.~F.~L.~Ward and Z.~Was, 
\href{https://journals.aps.org/prd/abstract/10.1103/PhysRevD.63.113009}{Phys.\ Rev.\ D {\bf 63}, 113009  (2001).}
\bbt{evt2} R.~G.~Ping {\it et al.},  
\href{https://iopscience.iop.org/article/10.1088/1674-1137/32/8/001}{Chin.\ Phys.\ C {\bf 32}, 599  (2008).}
\bbt{evt2-01} D.~J.~Lange, 
\href{https://www.sciencedirect.com/science/article/pii/S0168900201000894}{Nucl.\ Instrum.\ Meth.\ A {\bf 462},152 (2001).}
\bbt{geant4}  S.~Agostinelli {\it et al.} (GEANT4 Collaboration), 
\href{http://refhub.elsevier.com/S0370-2693(21)00497-4/bib65DADF2C57A74EB63401158AF79BAA84s1}{Nucl.\ Instrum.\  Methods Phys. Res., Sect. \ A {\bf 506}, 250  (2003).}
\bbt{geant4-01} J.~Allison {\it et al.}, 
\href{https://ieeexplore.ieee.org/document/1610988}{IEEE Trans.\ Nucl.\ Sci.\  {\bf 53}, 270  (2006).}
\bbt{XUM} M.~Xu  {\it et al.}, 
\href{https://iopscience.iop.org/article/10.1088/1674-1137/33/6/005}{Chin.\ Phys.\ C {\bf33}, (2009) 428.}
\bbt{PDG2020} P.~A.~Zyla {\it et al.} (Particle Data Group),  
\href{https://academic.oup.com/ptep/article/2020/8/083C01/5891211}{Prog. Theor. Exp. Phys. {\bf 2020}, 083C01 (2020).}
\bibitem{James:1975dr} F.~James and M.~Roos, 
\href{https://www.sciencedirect.com/science/article/abs/pii/0010465575900399?via%3Dihub}{Comput.\ Phys.\ Commun.\  {\bf 10}, 343 (1975).}
\bibitem{BESIII:2020nme}
M.~Ablikim {\it et al.} (BESIII Collaboration),
\href{https://iopscience.iop.org/article/10.1088/1674-1137/44/4/040001}{Chin. Phys. C \textbf{44}, 040001 (2020).}



 \end{thebibliography}
\end{document}